\documentclass[a4paper,USenglish, autoref,cleveref,thm-restate]{lipics-v2021}

\usepackage{tikz}

\usetikzlibrary{automata,arrows,positioning,calc}
\usetikzlibrary{patterns} 
\usepackage{thmtools}
\usepackage{todonotes}
\usepackage{amsmath}
\usepackage{amssymb}
\usepackage{float}
\usepackage{mathtools}
\numberwithin{equation}{section}
\usepackage{thmtools}
\usepackage{hyperref}
\usepackage{url}

\title{Base Fee Manipulation In Ethereum's EIP-1559 Transaction Fee Mechanism}  
\hideLIPIcs 
\author{Sarah Azouvi\footnote{\label{note1}The authors of this work are listed alphabetically.}}{Unaffiliated, United Kingdom}{s.azouvi@gmail.com}{0000-0002-7133-1937}{}
\author{Guy Goren}{Protocol Labs, Israel}{guy.goren@protocol.ai}{0000-0003-2158-161X}{}
\author{Lioba Heimbach}{ETH Zurich, Switzerland}{hlioba@ethz.ch}{0000-0002-8258-1712}{}
\author{Alexander Hicks}{UCL, United Kingdom}{alexander.hicks@ucl.ac.uk}{0000-0002-8941-9566}{}

\authorrunning{S. Azouvi, G. Goren, L. Heimbach and A. Hicks} 

\Copyright{Sarah Azouvi, Guy Goren, Lioba Heimbach, Alexander Hicks} 



\supplement{}
\supplementdetails[subcategory={Source Code}]{Software}{https://github.com/liobaheimbach/Base-Fee-Manipulation-In-Ethereum-s-EIP-1559-Transaction-Fee-Mechanism}

\acknowledgements{We thank Andrei Constantinescu for the helpful ideas and discussions.}

\nolinenumbers 

\EventEditors{Rotem Oshman}
\EventNoEds{1}
\EventLongTitle{37th International Symposium on Distributed Computing (DISC 2023)}
\EventShortTitle{DISC 2023}
\EventAcronym{DISC}
\EventYear{2023}
\EventDate{October 10-12, 2023}
\EventLocation{L'Aquila, Italy}
\EventLogo{}
\SeriesVolume{281}
\ArticleNo{11}

\begin{document}
\definecolor{LIGHTGREEN}{HTML}{96E6B3}
\definecolor{RED}{HTML}{DA3E52}

\definecolor{YELLOW}{HTML}{F8BE57}
\definecolor{VIOLET}{HTML}{7E6B8F}
\definecolor{LIGHTBLUE}{HTML}{A3D9FF}
\definecolor{DARKBLUE}{HTML}{06AED5}

\newcommand{\rcur}{r_{cur}}
\newcommand{\rpred}{r_{pred}}
\newcommand{\spred}{s_{pred}}
\newcommand{\starget}{s^*}
\newcommand{\btarget}{b^*}

\newcommand{\maxblocksize}{\textsf{maxBlockSize}}
\newcommand{\blocks}{\mathcal{B}}
\newcommand{\attackprofit}{\textsf{attackProfit}}
\newcommand{\attackloss}{\textsf{attackLoss}}
\newcommand{\maxfee}{\textsf{maxFee}}
\newcommand{\tx}{\textit{tx}}
\newcommand{\TFM}{\textsf{TFM}}
\newcommand{\miner}{x}

\newcommand\T[1]{\vspace{2pt}\noindent\textbf{#1}}
\newcommand\TA[1]{\vspace{0pt}\noindent\textbf{#1}}

\newcommand{\guyg}[1]{[\textcolor{blue}{\textbf{gg:}} {\footnotesize
\textcolor{purple}{#1}]}}
\setlength{\textfloatsep}{5pt} 
\setlength{\abovecaptionskip}{5pt}
\setlength{\belowcaptionskip}{5pt}
\interlinepenalty=-1
\setlength{\textfloatsep}{5pt}

\setlength{\abovedisplayskip}{3pt}
\setlength{\belowdisplayskip}{3pt}
\maketitle

\begin{abstract}
In 2021 Ethereum adjusted the transaction pricing mechanism by implementing EIP-1559, which introduces the \emph{base fee} --- a network fee that is burned and dynamically adjusts to the network demand. The authors of the Ethereum Improvement Proposal (EIP) noted that a miner with more than 50\% of the mining power could be incentivized to deviate from the honest mining strategy. Instead, such a miner could propose a series of empty blocks to artificially lower demand and increase her future rewards. In this paper, we generalize this attack and show that under rational player behavior, deviating from the honest strategy can be profitable for a miner with less than 50\% of the mining power. We show that even when miners do not collaborate, it is at times rational for smaller miners to join the attack. Finally, we propose a mitigation to address the identified vulnerability.
\end{abstract}

\section{Introduction}
\label{sec:intro}

Ethereum occupies a central role in the blockchain and decentralized application landscape. Not only does Ethereum's market capitalization currently exceed \$225 billion~\cite{coinmarketcap}, but it is also the leading platform for smart contracts and decentralized finance. Thus, Ethereum has revolutionized the way people envision and interact with blockchain technology.

The proposal of Ethereum Improvement Proposal \#1559 (EIP-1559)~\cite{eip1559} in April 2019 and its later deployment on Ethereum's mainnet in August 2021, mark a significant milestone for the Ethereum network. EIP-1559 reshaped Ethereum's transaction fee mechanism and remains in place to this day. One of the main goals of EIP-1559 is to simplify the bidding process by reducing the need for complex fee estimation algorithms while ensuring that the mechanism is \textit{incentive compatible} --- the best strategy for all players is to follow the protocol as intended. In particular, the mechanism should be both truthful and not incentivize miner or user bribes, as articulated by Roughgarden~\cite{roughgarden2020transaction}. EIP-1559 has been shown to be incentive compatible~\cite{roughgarden2020transaction,roughgarden2021transaction} when miners are assumed to be \textit{myopic}, i.e., short-sighted in the sense that they are maximizing only their immediate profits. Furthermore, an empirical study~\cite{10.1145/3548606.3559341} concludes that EIP-1559 succeeded in making fees easier to estimate and in reducing delays.

EIP-1559 represents a departure from the previous \textit{first-price auction} system, wherein users submit bids and pay the exact amount of their bid if their transaction was included. Importantly, the entirety of the fees paid by users is awarded to miners. This is no longer the case under EIP-1559 which introduced a base fee, a portion of the transaction fee that is burned and not awarded to miners. The base fee varies according to the fill rate of blocks --- a proxy for network demand. Blocks exceeding a predefined target size increase the base fee and blocks below this target size decrease the base fee. Miners are then compensated for creating blocks through a block reward and user tips, i.e., user fees exceeding the base fee.

Considering the substantial financial value being traded on the Ethereum network, it is highly probable that profitable and rational deviations will occur when possible. Thus, it is essential for EIP-1559 to be incentive compatible as the protocol guarantees rely on participants following the intended behavior. Deviations from the intended behavior are considered attacks on the systems. As we will see, the dynamic nature of the base fee opens the door for possible rational attacks by miners. Miners have control over the fill rate of blocks and may thus choose to mine emptier blocks in order to decrease the base fee and increase their future profits, i.e., tips paid on top of the base fee by users. This genre of attack strategies has been acknowledged in Ethereum's EIP-1559 proposal and has been explored under various assumptions in previous research works~\cite{roughgarden2020transaction,hougaard2022farsighted}.

This paper presents an analysis of the potential for minority attacks on Ethereum's EIP-1559 transaction fee mechanism under the conservative assumption of a steady demand curve. Our results show that the mechanism is vulnerable to such a 20\% minority attack.\footnote{An individual entity with more than 30\% staking power currently exists in the Ethereum network~\cite{grandjean2023ethereum}.} Additionally, we show that smaller miners may be incentivized to join in on the attack. We provide general insights into when deviating from the prescribed strategy is rational and note that, due to the nature of our model and assumptions, the results are applicable in a wide range of scenarios. We also explain how the attack can be initiated by an Ethereum user (rather than a miner), i.e., show the incentives of users to enact bribes. Finally, to address the identified vulnerability, we propose a mitigation and evaluate its effectiveness through simulations.

\section{Basic block reward mechanism in Ethereum}\label{sec:background}
\TA{Block proposals.}
Whether Ethereum's blockchain relies on proof-of-work (PoW) or proof-of-stake (PoS) as sybil resistance, it relies on a leader election to determine the proposer of the next block. To be precise, in a PoW blockchain, miners compete to solve a computational puzzle, and the likelihood of a miner being chosen is based on their share of the network's computational power. In a PoS blockchain, on the other hand, miners stake amounts of the blockchain's native currency to participate and are randomly selected to create a new block with probability proportional to the amount they stake. In both cases, the ideal process is memoryless so each leader election is independent of the previous one. Our model covers both PoW and PoS, hence, the results of this paper apply to both types of blockchains. 

When a miner is chosen to propose a block, they select a set of pending transactions to include in the next block and broadcast it to the network. Upon successful inclusion of their block in the blockchain, the miner receives two rewards: a fixed reward of newly minted currency (Ether in Ethereum's  case), and a variable reward from the transaction fees of the included transactions. The block reward is fixed, regardless of the block's content or the miner that proposes it, so our analysis focuses on the potential for miners to increase their revenue via transaction fees. Therefore, we do not consider the block reward.

\T{Transaction fees under EIP-1559.}\label{sec:Transaction fees}
Ethereum transactions involve a set of instructions that are carried out by miners when the transaction is added to the blockchain. To prevent users from overwhelming the network with bogus transactions, users must pay \textit{transaction fees}. These fees should reflect the amount of computational resources needed to execute the instructions, measured in units of \textit{gas} and priced in \textit{Gwei}. Note that 1 Gwei $\triangleq$  10$^{-9}$ Ether.

The EIP-1559 transaction fee includes a \textit{base fee}, which is paid per unit of gas and varies to balance the supply of gas with the demand for gas. To be exact, the base fee $b[i]$ for block $i$ is determined from the base fee $b[i-1]$ and size $s[i-1]$ of the previous block as follows:
\begin{align}\label{eq:basefee}
    b[i] \triangleq b[i-1] \cdot \left(1+ \phi \cdot\frac{s[i-1]-\starget}{\starget}\right).
\end{align}
Thus, the base fee is determined by comparing the size (measured in consumed gas units) of the previous block to a target block size $\starget$. If the block is larger than the target size, it indicates high demand for gas, and the base fee is increased to decrease demand. Conversely, if the block size is smaller than the target, it indicates low demand for gas, and the base fee is decreased to increase demand. The sensitivity of the base fee to the size of the previous block is determined by the adjustment parameter $\phi$ that is currently set to $\frac{1}{8}$ on Ethereum. We note that $b[0]$ was set to 1 Gwei initially and that the maximum valid block size is $2\starget\!$. 

When creating a transaction under the EIP-1559 mechanism, in addition to the gas limit~($g$), the user must specify the fee cap~($c$) which is the maximum fee per gas unit they are willing to pay, and a maximum tip per gas unit~($\varepsilon$) which is the priority fee. The transaction will be included in a block only if the fee cap is greater than or equal to the base fee~($b$). The total fee paid by the user is $\tilde{g} \cdot \min\{b+\varepsilon,c\}$, where $\tilde{g}<g$ is the actual gas consumed by the transaction.%
\footnote{The gas limit ($g$) specifies the amount of gas units available for the execution of the transaction. The amount of gas needed to execute a transaction is not always predictable in advance. Moreover, if the needed amount of gas exceeds~$g$, then $g$ gas units are consumed and paid for but the transaction fails.}
A portion of the fee $\tilde{g}\cdot b$ is burnt and the remaining $\tilde{g}\cdot\min\{\varepsilon, c-b\}$ goes to the miner as a tip. The EIP-1559 mechanism aims for users to bid small tips that only cover a miner's costs~\cite{eip1559,roughgarden2020transaction}. Miners are intended to include all transactions that have a fee cap greater or equal to the base fee and prioritize transactions with higher fees only if the maximal block size ($s_{\textit{max}}$) is exceeded. For simplicity, we assume that all transactions are of the same size and have a sufficient gas limit (i.e., $g=\tilde{g}=1$) to eliminate considerations including knapsack and gas estimation and keep the focus on the core matter. E.g., a large transaction is modeled by multiple smaller transactions.

\section{Model and Assumptions}\label{sec:model}
In the following, we present our model and outline our assumptions. We highlight that our model follows Roughgarden's~\cite{roughgarden2020transaction} very closely with the exception that we do not restrict miners to be myopic. Specifically, while Roughgarden~\cite{roughgarden2020transaction} considers only a miner's immediate profits (myopici), we will also consider her future profits.

\T{Users.} We assume that users are rational agents who want their submitted transactions to be processed and included in the blockchain. The cost of having one's transaction~(\tx{}) included in the blockchain under EIP-1559 depends on the base fee ($b$), fee cap ($c_\tx$), and the maximum tip ($\varepsilon_\tx$), which we described in Section~\ref{sec:Transaction fees}. Each transaction \tx{} has a value $v_\tx$ that is private to the user who proposes it, which can be thought of as the maximum price that the user is willing to pay for the transaction \tx{} to be included in the blockchain. We, therefore, take the utility of a user proposing \tx{} at each block to be $u(\tx) = v_\tx - \min\{b+\varepsilon_\tx, c_\tx\}$ if \tx{} is included and 0 otherwise. The user will then adjust $c_\tx$ and $\varepsilon_\tx$ according to her bidding strategy, while $b$ is determined by the size of the past blocks according to Equation~\ref{eq:basefee}.

\T{Miners.} Miners produce blocks that include transactions to be executed. In our model, the miner to propose the next block is chosen at random. Drawings constitute independent experiments. In each drawing, a miner~$X$ is chosen with probability~$p_\miner$, where~$p_\miner$ is the miner's share of the total network power. We assume that the power distribution in the network changes slowly, hence, to simplify the analysis we model the system as having a fixed network power distribution, i.e., miners' network power does not change with time. 

A miner has dictatorial power over which transactions to include in the block she produces. We assume that miners are rational agents who wish to maximize their profit and therefore choose which transactions to include in their block according to a strategy that maximizes their profit. Aside from their transaction picking strategy, we assume that miners behave ``honestly'' -- that is, as specified by the protocol. We highlight that by not considering other forms of deviating from the protocol, we strengthen our result, showing that even ``practically honest'' miners would deviate from EIP-1559.

\T{Collaboration.} We adopt a standard buyers-sellers perspective. We assume that miners do not collaborate with each other, as they can be viewed as one miner with more power. Similarly, we assume that users do not collaborate with one another since they are competing for the same scarce resource. However, a user (buyer) and a miner (seller) can communicate and adjust their strategies for mutual benefit.
In particular, a (sophisticated%
\footnote{It is expected that sophisticated participants will emerge, actively seeking opportunities to generate excess profits (reduce costs). This expectation is supported by the history of traditional exchange systems, where a multitude of players specialize in high-frequency trading.})
user and a miner will collaborate if it benefits both of them. That is, the user pays fewer fees for her transaction while the miner receives more fees. The collaboration is thus rational behavior, leading to improved outcomes for both.

\T{Steady State.} The distribution of transaction values is represented by a demand curve, and the standard demand curve is a decreasing function; the higher the fee is, the fewer transactions are willing to pay it. In our work, we make no assumptions about the shape of the demand curve other than that it is a decreasing function. We consider a system in \textit{steady state}, which we define as a system in which the demand curve does not change over time, i.e., it is the same whenever a miner creates a block. This steady state assumption is good for several reasons: (1) it keeps us on par with Roughgarden's work~\cite{roughgarden2020transaction}, (2) it simplifies the analysis by removing noisy components that can obscure the core principles, and (3) it is a conservative assumption that strengthen our results in comparison to others. For example, the ``steady influx'' model --- where there is a fixed influx of transactions to the network --- assumes that miners are able to manipulate the demand curve in their favor, i.e., artificially increase demand by delaying the inclusion of transactions. Thus, when the attack strategy we will present in Section~\ref{sec:attack} is beneficial under the \textit{steady state} it will be at least as beneficial under the ``steady influx'' model, but not vice versa. Hence, our steady demand curve assumption makes our results more general and robust, as it applies to a wider range of adversarial assumptions.

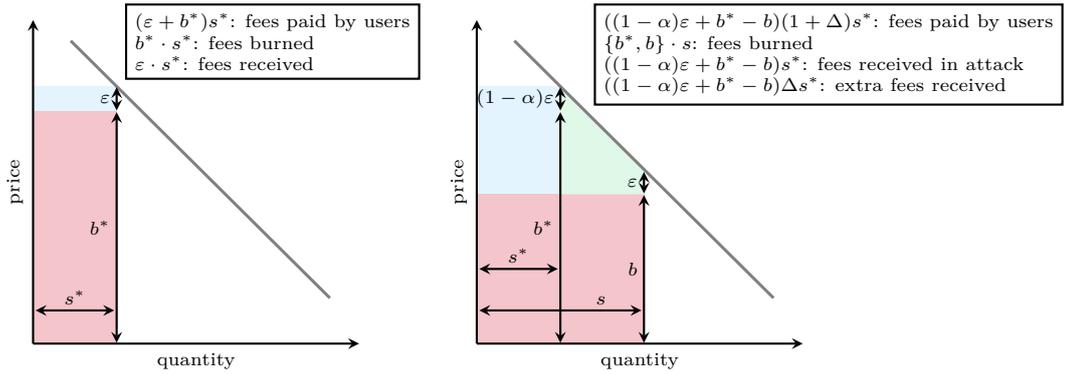
\begin{figure}[t]
\centering
\begin{subfigure}[t]{0.395\linewidth}
\centering
\begin{tikzpicture}[scale=1.1,font=\scriptsize]
    \node [below right, draw, align = left, thick] at (1.1, 4.1) {
    $(\varepsilon+\btarget) \starget$: fees paid by users \\    
    $\btarget\cdot  \starget$: fees burned \\
    $\varepsilon\cdot  \starget$: fees received
    };
       \fill [LIGHTBLUE!30]
        (0,3.1)
       -- (1,3.1)
       -- (1,2.8)
       -- (0,2.8)
       -- cycle;
     
       \fill [RED!30]
        (0,2.8)
       -- (1,2.8)
       -- (1,0)
       -- (0,0)
       -- cycle;
        \draw[-stealth, thick] (0, 0) -- (3.9, 0) node[midway, below,sloped] {quantity};
        \draw[-stealth, thick] (0, 0) -- (0, 3.9) node[midway, above,sloped] {price};
        \draw[scale=1,line width=0.4mm, domain=0.45:3.55, smooth, variable=\x, gray] plot ({\x}, {3.5-\x+0.6});

        \draw[thick,stealth-stealth] (1,2.79) -- (1,0.01) node[midway,left,inner sep =2pt] {$\btarget$};
        \draw[thick,stealth-stealth] (1,3.09) -- (1,2.81) node[pos = 0.5,left,inner sep =2pt] {$\varepsilon$};
   
        \draw[thick,stealth-stealth] (0.03,0.4) -- (0.97,0.4) node[midway,above,inner sep =2pt] {$\starget$};

    \end{tikzpicture}\vspace{-6pt}
        \caption{Sample steady state demand curve, $\starget$ transactions are willing to pay $\btarget +\varepsilon$, where $\btarget$ is the base fee that corresponds to the target size. The red area is the amount of fees being burned, whereas the blue area are the fees received by the miner.}
        \label{fig:demand_curve}\vspace{-6pt}
    \end{subfigure}\hfill
    \begin{subfigure}[t]{0.583\linewidth}
\centering
\begin{tikzpicture}[scale=1.1,font=\scriptsize]

    \node [below right, draw, align = left, thick] at (1.4, 4.1) {
    $((1-\alpha)\varepsilon+ \btarget -b) (1+\Delta) \starget $: fees paid by users\\
    $\{\btarget,b\}\cdot  s$: fees burned \\
    $((1-\alpha)\varepsilon+ \btarget -b)  \starget$: fees received in attack\\
    $((1-\alpha)\varepsilon+ \btarget -b) \Delta \starget $: extra fees received
    };

       \fill [LIGHTBLUE!30]
        (0,3.1)
       -- (1,3.1)
       -- (1,1.8)
       -- (0,1.8)
       -- cycle;
       \fill [RED!30]
        (0,1.8)
       -- (2,1.8)
       -- (2,0)
       -- (0,0)
       -- cycle;
       \fill [LIGHTGREEN!30]
        (2,2.1)
       -- (2,1.8)
       -- (1,1.8)
       -- (1,3.1)
       -- cycle;
        \draw[-stealth, thick] (0, 0) -- (3.9, 0) node[midway, below,sloped] {quantity};
        \draw[-stealth, thick] (0, 0) -- (0, 3.9) node[midway, above,sloped] {price};
        \draw[scale=1,line width=0.4mm, domain=0.45:3.55, smooth, variable=\x, gray] plot ({\x}, {3.5-\x+0.6});

        \draw[thick,stealth-stealth] (1,2.79) -- (1,0.01) node[midway,left,inner sep =2pt] {$\btarget$};
        \draw[thick,stealth-stealth] (1,3.09) -- (1,2.81) node[pos = 0.5,left,inner sep =2pt] {$(1-\alpha)\varepsilon$};
        \draw[thick,stealth-stealth] (2,1.79) -- (2,0.01) node[midway,left,inner sep =2pt] {$b$};
        \draw[thick,stealth-stealth] (2,2.07) -- (2,1.81) node[pos = 0.5,left,inner sep =2pt] {$\varepsilon$};
        \draw[thick,stealth-stealth] (0.03,0.9) -- (0.97,0.9) node[midway,above,inner sep =2pt] {$\starget$};
        \draw[thick,stealth-stealth] (0.03,0.4) -- (1.97,0.4) node[pos =0.75,above,inner sep =2pt] {$s$};       
    \end{tikzpicture}\vspace{-6pt}
        \caption{Sample steady state demand curve with lowered base fee $b$. If the base fee was lowered to $b<\btarget$, the demand is increased to $s\geq\starget$. An honest miner will fill up the block with all transactions, paying at least the base fee. The red area indicates the amount of fees being burned, the blue area the fees the attacking miner~$X$ receives, and the green area is the additional fees received by the honest miner~$Y$ for including all transactions. The scaling factor $\Delta$ dictates the size of the green area.}
        \label{fig:demand_curve_honest}\vspace{-6pt}
\end{subfigure}
\caption{Example demand curves for Ethereum transactions. The quantity ($x$-axis) indicates the number of transactions willing to pay the transaction fee (price shown $y$-axis).}
\end{figure}

The desired dynamics of EIP-1559 in the steady state are that produced blocks are of target size $\starget$ and that the base fee remains constant at what we refer to as the target base fee $\btarget$, i.e., $s[i]=\starget$ and $b[i]=\btarget$ for all~$i$. Further, EIP-1559's desired bidding dynamics are for the user's optimal strategy to be honest in reporting the value of a transaction via~$c_\tx$ and to offer a minimal tip. In other words, in the steady state, EIP-1559 should lead to a Nash-equilibrium with the following strategies: (users) honest value-reporting, and (miners) including all $\tx$s with $c_\tx>\btarget$. Then a block produced during steady state would include $\starget$ transactions that are each paying $\btarget+\varepsilon$ in fees. Thus, $\btarget \cdot \starget$ is burned (\textcolor{RED}{red area} in Figure~\ref{fig:demand_curve}), and $\varepsilon \cdot \starget$ is received by the miner (\textcolor{LIGHTBLUE}{blue area} in Figure~\ref{fig:demand_curve}).

\section{A Miner's Deviation from the Honest Strategy}\label{sec:attack}
In the following, we consider a miner~$X$ who controls a proportion $p_x$ of the network's mining power, i.e., the probability that $X$ proposes the next block is $p_x$. 

\T{Honest strategy.} The honest strategy for $X$ is always to include the maximum possible number of transactions whose gas fee covers at least the base fee. As we consider a system in steady state, the demand curve does not change over time. Further, the honest user bids $c_\tx =\btarget+\varepsilon $ and $\varepsilon_\tx = \varepsilon$. Thus, miner~$X$ will always propose a block that is exactly the target size $\starget$. The payout received by miner~$X$ for every proposed block is, therefore, $\starget\cdot \varepsilon$, where $\varepsilon$ is the tip the miner receives. The costs for the users are $(\btarget+\varepsilon)\starget$.

\T{Deviation from honest strategy.} We continue by outlining a strategy miner~$X$ and sophisticated users can employ to manipulate the base fee which results in increasing $X$'s profit and reducing users' costs. When $X$ proposes a block, for which the preceding block was not created by $X$, she proposes an empty block to reduce the base fee $b$ for subsequent blocks. The miner will receive no payout for this block. Any other consecutive blocks $X$ is chosen to propose, she will propose at target size $\starget$, profiting from the difference between the targeted and the reduced base fee. To highlight the miner's profit, we assume in the following that users are motivated by slightly reducing their costs. Specifically, users continue to submit transactions with $c_\tx =\btarget+(1-\alpha)\varepsilon $ and $\varepsilon_\tx = \phi\cdot \btarget+(1-\alpha)\varepsilon$, where $0 < \alpha \leq 1$.  
Thus, the miner will receive at least the difference in base fees, i.e., $ \phi\cdot\btarget$, and the user will pay $\alpha\cdot\varepsilon$ less for her transaction. By making this assumption, we let the miner extract most of the excess profit. In Section~\ref{sec:user} we describe the complementary attack where most of the excess value is gained by the users.

Whenever $X$ proposes a $\starget$ sized block with an artificially reduced base fee, $X$ receives
\begin{equation}\label{eq:xreward}
     \starget \left( \phi\cdot \btarget +(1-\alpha)\varepsilon \right),
\end{equation}
where $\phi\cdot \btarget$ is the base fee reduction and $\alpha$ represents the proportional reduction of the tip paid by the users. For simplicity, we assume the attack finishes whenever miner~$X$'s consecutive turns as proposer finishes. The deviation can only become more profitable if $X$ can continue the attack after an honest proposer, thus, our results apply without the simplification and their generality is not reduced.

In Theorem~\ref{thm:oneminer} we compute the expected reward of $X$ following the honest strategy, as well as the aforementioned described deviation from the honest strategy. By comparing the payout of consecutive turns of $X$ in both strategies, we find that it is rational also for a miner with less than 50\% of the power to deviate from the honest strategy.
\begin{restatable}{thm}{oneminer}
\label{thm:oneminer}
In expectation, it is rational for miner~$X$ to deviate from the honest strategy,~if 
    \begin{equation*}
    p_x > \frac{ \varepsilon}{\phi\cdot \btarget +(1-\alpha)\varepsilon}.
    \end{equation*}
\end{restatable}
\begin{proof}
See Appendix~\ref{app:proofs}.
\end{proof}

\begin{figure}[th]
\centering
\begin{subfigure}[t]{0.315\linewidth}
\centering
\includegraphics[scale =1]{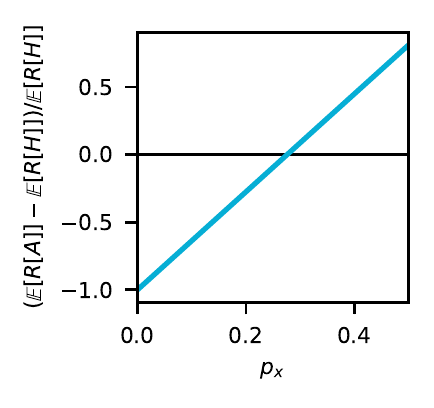}\vspace{-6pt}
\caption{Relative difference shown as a function of $p_x$. We set $\alpha = 0.5$, $\varepsilon/\btarget =1/ 25$.} \label{fig:px}\vspace{-6pt}
\end{subfigure}   
\hfill
\begin{subfigure}[t]{0.31\linewidth}
\centering
\includegraphics[scale =1]{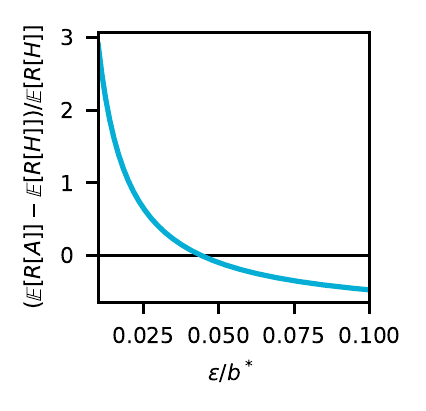}\vspace{-6pt}
\caption{Relative difference shown as a function of $\varepsilon/\btarget$. We set $p_x =0.3$, $\alpha = 0.5$.} \label{fig:ratiodb}\vspace{-6pt}
\end{subfigure}   
\hfill
\begin{subfigure}[t]{0.315\linewidth}
\centering
\includegraphics[scale =1]{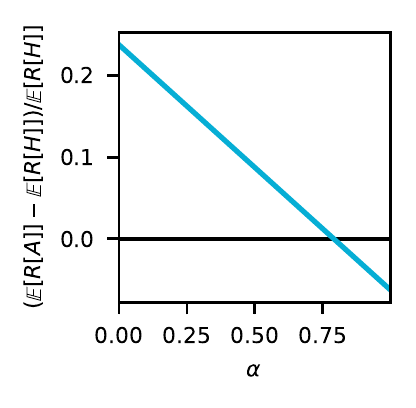}\vspace{-6pt}
\caption{Relative difference shown as a function of $\alpha$. We set $p_x =0.3$, $\varepsilon /\btarget = 25$.} \label{fig:a}\vspace{-6pt}
\end{subfigure}   
\caption{Relative difference between the expected reward of attack and honest strategy as a function of $p_x$ (cf. Figure~\ref{fig:px}), $\varepsilon/\btarget$ (cf. Figure~\ref{fig:ratiodb}) and $\alpha$ (cf. Figure~\ref{fig:a}). We set $\phi= 1/8$, as set in Ethereum. It is rational for $X$ to perform the attack whenever the relative difference is positive. The equations to produce these graphs can be found in the proof of Theorem~\ref{thm:oneminer} in Appendix~\ref{app:proofs}.
}\label{fig:diff}\vspace{-4pt}
\end{figure}

To better illustrate, when it is rational behavior for miner~$X$ to deviate from the honest strategy, we plot the relative difference between the expected reward of the attack and the honest strategy in Figure~\ref{fig:diff}. For many realistic parameter configurations, it is rational behavior for miner~$X$ to perform the attack and thereby manipulate the base fee. In Figure~\ref{fig:px}, we plot the profitability of the attack in comparison to the honest strategy dependent on $X$'s mining power $p_x$. We set $\phi=1/8, \varepsilon/\btarget=1/25$, and $\alpha = 0.5$. Notice that even miners with a mining power of less than $0.3$ are expected to profit from executing the attack. To underline the expected profitability of the attack, even for small miners, we vary the ratio between the tips and the base fee ($\varepsilon/\btarget$) in Figure~\ref{fig:ratiodb} and set $p_x = 0.3$. Additionally, we vary $\alpha$, the share of the tips the users keep for themselves, in Figure~\ref{fig:a}, and again set $p_x = 0.3$. We conclude that there are multiple realistic parameter configurations for which the outlined attack is profitable, even for a miner with less than 50\% of the mining power. Thus, as individual Ethereum pools control in excess of 30\% of the staking power~\cite{grandjean2023ethereum} such an attack is realistic.

\section{The Attack's Effect on Other Miners}\label{sec:otherMiners}
In the following, we consider a scenario where a miner~$X$ with staking power $p_x$ ($ 0< p_x < 1$) exists for which it is rational behavior to perform the base fee manipulation studied in Section~\ref{sec:attack}. We then analyze the effect of a miner~$Y$ with staking power $p_y$, where $ 0< p_y < p_x$, observing that $X$ performs the base fee manipulation attack. More precisely, we study whether it is rational behavior for $Y$ to join the attack partially, i.e., $Y$ will always propose blocks at target size and thereby help keep the base fee artificially low in Section~\ref{sec:keeplow}. Additionally, we will also study when $Y$ would rationally join the attack entirely, i.e., $Y$ also proposes empty blocks when the base fee is not already artificially lowered in Section~\ref{sec:start}. We remark that, throughout, we always assume that miners $X$ and $Y$ do not collaborate.

\subsection{Joining the Attack} \label{sec:keeplow}
Consider a miner~$Y$ that observes $X$ performing the attack outlined from Section~\ref{sec:attack}. Miner~$Y$ is selected as the proposer for a block that follows $X$'s turn as proposer, i.e., the base fee is currently artificially lowered to $(1-\phi)\btarget$. We analyze whether it is rational for $Y$ to follow the honest strategy, i.e., propose the largest block possible and thereby increase the base fee again, or to join the attack and continue keeping the base fee artificially low.

\T{Honest strategy.} We first describe the honest strategy. When it is miner~$Y$'s turn to propose a block at an artificially lower base fee, miner~$Y$ proposes a block with the most transactions possible. Note that the number of transactions is restricted both by the demand at the current base fee ($b$), as well as the maximum block size, which is twice the target size ($2\starget$). We now consider the demand curve that is drawn in Figure~\ref{fig:demand_curve_honest}. The demand at price $\btarget+\varepsilon$, where $\btarget$ is the target base fee price and $\varepsilon$ the tip, corresponds to a block of target size $\starget$. Miner~$Y$ will propose a block with the artificially lowered base fee $b = (1-\phi) \btarget$. The demand at this new price $b+\varepsilon$ is represented as $s$ in Figure~\ref{fig:demand_curve_honest}. Recall, that we make no assumptions about the shape of the demand curve. To make our results stronger, we consider the best case for the honest strategy. Namely, due to the increased demand, $Y$ can propose a block of maximum size (and reap the resulting extra rewards). That is, $s =2\starget$, after $X$'s turn. Thus, the payout for miner~$Y$ proposing a block of size $2 \starget$ is given by 
\begin{equation}
\starget \left( \phi\cdot \btarget +(1-\alpha)\varepsilon \right) (1+\Delta),
\end{equation}
where $\starget \left( \phi\cdot \btarget +(1-\alpha)\varepsilon \right)$ is the payout the attacking miner~$X$ would receive (cf. Equation~\ref{eq:xreward}) and $\Delta\in[0,1]$ is a scaling factor that dictates how much additional rewards miner~$Y$ received for mining a maximum size block. While $\Delta =0$ would indicate that $Y$ earns exactly as much as $X$, i.e., all extra transactions are capped at exactly the base fee, $\Delta =1$ would indicate that miner~$Y$ earns twice the rewards of $X$, i.e., all extra transactions are capped at the highest possible price of~$\btarget+\varepsilon$.

After miner~$Y$ proposes the block, she will be chosen to propose the next block with a probability of $p_y$. We continue with the approximation from before, i.e., the effect of a full block after an empty one leads approximately to the same point on the demand curve -- $(\starget,\btarget+\varepsilon)$. We note that this approximation is accurate up to a term in $O(\phi^2\btarget)$ where $b\in \Theta(\phi\btarget)$. Thus, miner~$Y$ proposes a block of size $\starget$ and is awarded $\starget\cdot \varepsilon$. From thereon out, she will continue doing so until her consecutive turns as a proposer stops.

With probability $p_x$, miner~$X$ will interrupt $Y$'s turn as proposer. Miner~$X$ will start the attack again and propose an empty block to lower the base fee. Note that our approximation of the base fee returning to steady-state levels is the best case for $Y$'s honest strategy and, thus, makes our results stronger. For all consecutive blocks proposed by $X$, miner~$X$ will propose target size blocks to keep the base fee~$b$ (i.e., $(1-\phi)\btarget$). If miner~$Y$ is again selected as a proposer after $X$'s turn, miner~$Y$ will proceed with her previously outlined strategy. 

At any point, with a probability of $1-p_y-p_x$, the consecutive turn of the two miners finishes. Note that for $Y$'s honest strategy analysis, we are only interested in these consecutive turns of the two miners as proposers, as we analyze the expected payout of the same sequences for the deviation from the honest strategy which we outline in the following. 

\T{Deviation from honest strategy.} We now describe a strategy $Y$ can employ to join the attack she observes. When it is $Y$'s turn to propose a block and the base fee is artificially lowered by $X$, miner~$Y$ will propose a block at the target size. Equivalently to the payout received by miner~$X$ for such a block (cf. Section~\ref{sec:attack}), miner~$Y$'s reward for proposing the block is given by  
\begin{equation}
\starget \left( \phi \cdot \btarget +(1-\alpha)\varepsilon \right).
\end{equation}

Following the block proposed by $Y$, miner~$Y$ is again selected to propose a block with probability $p_y$ and miner~$X$ with probability $p_x$. As long as the two miners have an uninterrupted sequence of block proposals, they both keep the base fee artificially low by always proposing target size blocks. Once their consecutive turn as proposers ends, we consider the attack finished. In Theorem~\ref{thm:partialattack}, we show that it can be profitable for such a miner~$Y$ with a mining power $p_y$ to join the attack, i.e., keep the base fee artificially low if she sees a miner~$X$ with a mining power $p_x>p_y$ perform the attack. Importantly, this is without assuming collaboration between the two miners. 
\begin{restatable}{thm}{partialattack}
\label{thm:partialattack}
In expectation, it is rational for a miner~$Y$ to deviate from the honest strategy and join $X$ in keeping the base fee low, if 
    \begin{equation*}
    p_y >\frac{\Delta ((1-\alpha)\varepsilon + \phi\cdot \btarget)}{(1-\alpha)\Delta\cdot\varepsilon +(1+\Delta) \phi\cdot \btarget - \alpha\cdot\varepsilon}.
    \end{equation*}
\end{restatable}
\begin{proof}
See Appendix~\ref{app:proofs}.
\end{proof}

\begin{figure}[t]
\centering
\begin{subfigure}[t]{0.327\linewidth}
\centering
\includegraphics[scale =1]{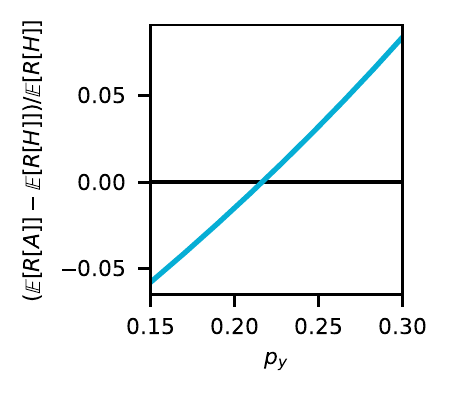}\vspace{-6pt}
\caption{Relative difference shown as a function of $p_y$. We set $p_x = 0.3$, $\Delta = 0.2$.} \label{fig:val_y_py}\vspace{-6pt}
\end{subfigure}   
\hfill
\begin{subfigure}[t]{0.322\linewidth}
\centering
\includegraphics[scale =1]{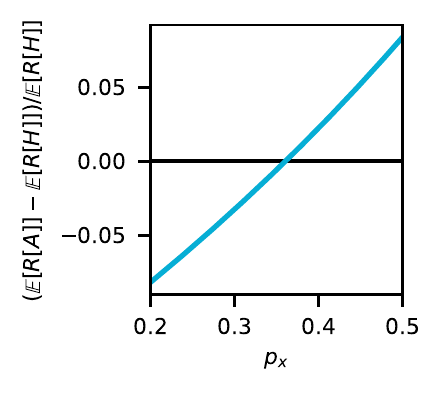}\vspace{-6pt}
\caption{Relative difference shown as a function of $p_x$. We set $p_y =0.6p_x$, $\Delta = 0.2$.} \label{fig:val_y_px}\vspace{-6pt}
\end{subfigure}   
\hfill
\begin{subfigure}[t]{0.312\linewidth}
\centering
\includegraphics[scale =1]{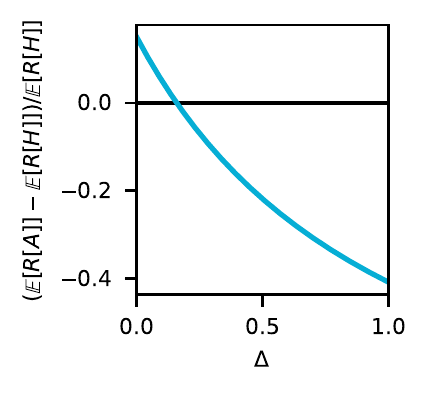}\vspace{-6pt}
\caption{Relative difference shown as a function of $\Delta$. We set $p_x =0.3$, $p_y = 0.18$.
} \label{fig:val_y_Delta}\vspace{-6pt}
\end{subfigure}   

\caption{Relative difference between the expected reward of attack and honest strategy as a function of $p_y$ (cf. Figure~\ref{fig:val_y_py}), $p_x$ (cf. Figure~\ref{fig:val_y_px}) and $\Delta$ (cf. Figure~\ref{fig:val_y_Delta}). We set $\phi= 1/8$, as implemented in Ethereum, $\varepsilon/\btarget= 1/25$, $\alpha = 0.5$ in all plots. It is rational for $Y$ to join the attack whenever the relative difference is positive. The equations to produce these graphs can be found in the proof of Theorem~\ref{thm:partialattack} in Appendix~\ref{app:proofs}.
}\label{fig:val_y_diff}\vspace{-2pt}
\end{figure}

To better gauge when it is profitable for miner~$Y$ to join the attack, we plot the relative difference between the expected return from the outlined attack and the expected return from the honest strategy in Figure~\ref{fig:val_y_diff}. We vary the mining powers of miner~$Y$ (cf. Figure~\ref{fig:val_y_py}) and miner~$X$ (cf. Figure~\ref{fig:val_y_px}), as well as $\Delta$ (cf. Figure~\ref{fig:val_y_Delta}). In all three plots we set $\phi=1/8, \varepsilon/\btarget=1/25$, and $\alpha = 0.5$. In Figure~\ref{fig:val_y_py}, we set $\Delta = 0.2$ and notice that even a miner~$Y$ with a mining power slightly larger than $0.2$ would be inclined to join the attacking miner~$X$ with a mining power of $0.3$ in manipulating the base fee. We observe in Figure~\ref{fig:val_y_px} that a miner~$Y$ that is only six-tenths of the size of miner~$X$ would also join the attack even if miner~$X$ only controls less than 40\% of the mining power. Finally, in Figure~\ref{fig:val_y_Delta}, we show the dependency of the attack's profitability for miner~$Y$ as a function of $\Delta$, i.e., the additional payout received by miner~$Y$ following the honest strategy when proposing a full block after the attack by $X$. Notice that while it is rational for a miner~$Y$ to join the attack for small $\Delta$'s, this is not the case for larger $\Delta$'s. We remark that this is due to the rewards from the full block mined if $Y$ follows the honest strategy being very significant for a large $\Delta$. Nevertheless, our results show that for realistic parameter configurations, it is rational for a miner~$Y$ to join the attack she sees a larger miner~$X$ perform --- even without assuming collaboration between the two.

\subsection{Join and Initiate the Attack}\label{sec:start}
In addition to only joining the attack, it is also possible for miner~$Y$, observing $X$ continuously performing the base fee manipulation, to also propose an empty block whenever she proposes a block with the target base fee ($\btarget$), knowing that $X$ will aid her in keeping the base fee low subsequently. We analyze, in the following, when it is more profitable for miner~$Y$ to join $X$'s attack in her entirety, as opposed to remaining honest.

\T{Honest strategy.} The honest strategy for miner~$Y$ is identical to that described in Section~\ref{sec:keeplow}. However, now it is also important to mention that anytime miner~$Y$ proposes a block with base fee $\btarget$, i.e., whenever $Y$ proposes a block that does not follow $X$'s attack, $Y$ will propose a target size ($\starget$) block. For proposing such a block, miner~$Y$ will receive $ \starget \cdot \varepsilon$.

\T{Deviation from honest strategy.} In the deviation from the honest strategy, $Y$ will propose an empty block whenever she mines a block where the base fee corresponds to the target base fee $\btarget$. Then with probability $p_y$, miner~$Y$ will also propose the next block and will profit from the difference between the base fee and the target base fee by mining a target size block. On the other hand, with probability $p_x$, miner~$X$ will propose the next block and will also mine a target size block -- keeping the base fee low. With Theorem~\ref{thm:entireattack}, we show that it can even be profitable for a miner~$Y$ to commence the attack if she knows that a larger miner~$X$ will help her in keeping the base fee artificially low. 
\begin{restatable}{thm}{entireattack}
\label{thm:entireattack}
    In expectation, it is rational for a miner~$Y$ to deviate from the honest strategy and lower the base fee, if 
$$
p_y > \frac{\varepsilon (1-p_x)}{(1-\alpha) \varepsilon + \phi \cdot \btarget (1-p_x) + (1+\Delta)\varepsilon\alpha p_x - \Delta p_x(\varepsilon+\phi\cdot \btarget)}.
$$\end{restatable}
\begin{proof}
See Appendix~\ref{app:proofs}.
\end{proof}

\begin{figure}[ht]
\centering
\begin{subfigure}[t]{0.317\linewidth}
\centering
\includegraphics[scale =1]{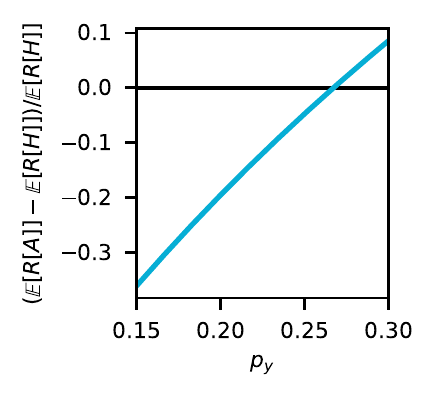}\vspace{-6pt}
\caption{Relative difference shown as a function of $p_y$. We set $p_x=0.3$, $\Delta=0.2$.} \label{fig:val_y_2_py}\vspace{-6pt}
\end{subfigure}   
\hfill
\begin{subfigure}[t]{0.312\linewidth}
\centering
\includegraphics[scale =1]{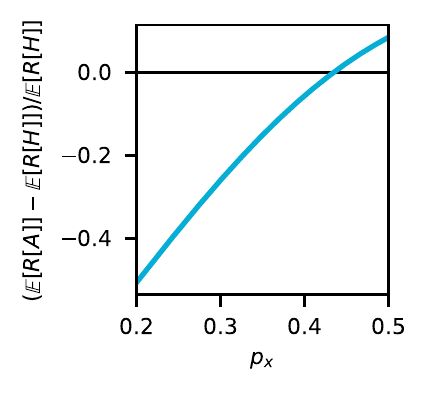}\vspace{-6pt}
\caption{Relative difference shown as a function of $p_x$. We set $p_y=0.6p_x$, $\Delta = 0.2$.
} \label{fig:val_y_2_px}\vspace{-6pt}
\end{subfigure}   
\hfill
\begin{subfigure}[t]{0.322\linewidth}
\centering
\includegraphics[scale =1]{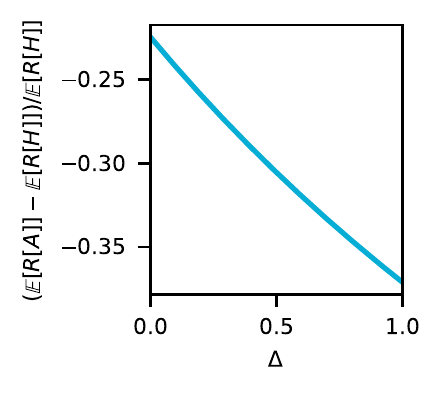}\vspace{-6pt}
\caption{Relative difference shown as a function of $\Delta$. We set $p_x =0.3$, , $p_y = 0.18$.} \vspace{-6pt}\label{fig:val_y_2_Delta}
\end{subfigure}   

\caption{Relative difference between the expected reward of attack and honest strategy as a function of $p_y$ (cf. Figure~\ref{fig:val_y_2_py}), $p_x$ (cf. Figure~\ref{fig:val_y_2_px}) and $\Delta$ (cf. Figure~\ref{fig:val_y_2_Delta}). We set $\phi= 1/8$, as implemented in Ethereum, $\varepsilon/\btarget= 1/25$, $\alpha = 0.5$ in all plots. It is rational for $Y$ to initiate new attacks in addition to $X$ whenever the relative difference is positive. The equations to produce these graphs can be found in the proof of Theorem~\ref{thm:entireattack} in Appendix~\ref{app:proofs}.
}\label{fig:val_y_2_diff}\vspace{-4pt}
\end{figure}

To better understand the parameter configuration under which it would be rational for $Y$ to also initiate the attack, we plot the relative difference of the reward of the attack in comparison to the reward of the honest strategy in Figure~\ref{fig:val_y_2_diff}. Note that we again set  $\phi=1/8, \varepsilon/\btarget=1/25$, and $\alpha = 0.5$. Figure~\ref{fig:val_y_2_py}, where we also set $p_x=0.3$ and $\Delta = 0.2$, shows that it can be profitable for a miner~$Y$ to initiate the attack, i.e., mine an empty block to lower the base fee, knowing that miner~$X$ will support her in keeping the base fee artificially low. Notice though that the threshold where it is rational for $Y$ to also start the attack is reached later in comparison to the threshold where it is rational for miner~$Y$ to only join the attack (cf. Figure~\ref{fig:val_y_py}). A similar picture paints itself when we look at the profitability of initiating the attack for $Y$ as a function of $X$'s mining power in Figure~\ref{fig:val_y_2_py}. Again we see that for a miner~$Y$, it can be profitable to start the attack only with the knowledge that $X$ will aid her in keeping the base fee low. Finally, in Figure~\ref{fig:val_y_2_Delta}, we show the relative difference between the expected profit of the outlined attack for $Y$ and the honest strategy. For the chosen parameter configuration, $p_x = 0.3$ and $p_y = 0.18$, the attack would actually never be profitable regardless of $\Delta$. We, thus, summarize that while it is only rational for $Y$ to initiate the attack for a reduced set of parameters, it is remarkable that this is even the case. By starting the attack, $Y$ is taking a loss for a larger miner~$X$ just based on the knowledge that she will be supported by $X$ in keeping the base fee low. Astonishingly, the collaboration of the two miners is not required.

\section{Possible Mitigations}\label{sec:mitigation}
To mitigate the attacks described in Sections~\ref{sec:attack} and~\ref{sec:otherMiners}, we focus on addressing the deviation of player $X$, as if $X$ does not deviate from the honest strategy, and neither does $Y$. We start by examining the trivial mitigation of reducing~$\phi$. Theorem~\ref{thm:oneminer} shows that decreasing $\phi$ by a factor of $\beta$ will require that $p_x$ is approximately $\beta$ times larger for a deviation to be profitable. For example, if Ethereum were to decrease its current $\phi$ of $1/8$ to $1/16$, it would require $p_x$ to be approximately twice as large for a deviation to be profitable. This approach can be effective during stable periods, but it might not be able to adjust quickly enough to changes in demand. Further, Leonardos et al.~\cite{10.1145/3479722.3480993} show how the value of~$\phi$ determines the trade-off between a base fee that adjusts too quickly (which can lead to chaotic behavior) and one that adjusts too slowly to fulfill its purpose.

The following question appears in EIP1559 FAQ~\cite{EIP1559FAQ}: \textit{``Won't miners have the incentive to collude to push down the base fee by making all their blocks less than half full?''} In response to the question Buterin proposes this mitigation: \textit{``Divert half of the collected base fees, that would otherwise be burned, into a special pool. Whenever a miner mines a new block add to her block reward a $1/8192$ portion of the amount in that pool. This will incentivize miners to maintain a higher base fee.''} One might falsely presume that this solution, to the above-posed question, might also be used to solve the deviation we outline in Section~\ref{sec:attack}. However, the two attacks are inherently different as the one we outline does not require miners to collude. Thus, this proposal does not solve our deviation. The added cost of the attack, i.e., the lost revenue from half of the base fee reduction, is distributed among many while the attacker's expected revenue remains unchanged. In the long run, this proposal only minimally reduces the attacker's expected revenue by $\btarget\cdot 2^{-13}$, which is typically significantly less than the expected rewards and, therefore, ineffective.

Another straw man to consider is to use the average of the previous $W$ block sizes instead of just the size of the last block to measure demand in the base fee update rule. This method may appear to be promising because it reduces the effect of an empty block on the following block by a factor of $W$, and its effect on the rate of adjustment (embodied in~$\phi$) is easily accounted for, adding an adjustment delay within $O(W)$. However, this mitigation fails to mitigate the attack, as it increases the opportunity for $X$ to profit from later blocks that are within a $W$ distance from the empty block, thereby increasing the expected profitability of the deviation. For example, if we use a two-block window (i.e., $W=2$) and $\phi=1/8$, the base fee is reduced by a smaller factor of only $\phi/W = -1/16$ to $b_1 = 15\btarget/16$ as desired. Nevertheless, even if the ensuing block is completely full (i.e., $s_1=2\starget$), the base fee for the block following it does not increase and remains $b_2 = b_1 = 15\btarget/16$ which means an additional profit opportunity for the deviation that compensates $X$ for the reduced factor. As a result, the deviation is not mitigated and is actually exacerbated.

We propose the following mitigation, to use a geometric sequence as weights to average the history of block sizes. Formally, for $q\in (0,1)$ denote
\begin{equation}
\label{eq:Savg}
\begin{split}
    s_{\textit{avg}}[i] & \triangleq \frac{1-q}{q}\sum_{k=1}^{\infty} q^k\cdot s[i-k+1]= (1-q)\cdot s[i] + q\cdot s_{\textit{avg}}[i-1],
\end{split}
\end{equation}
and replace $s[i]$ in Equation~\ref{eq:basefee} by $s_{\textit{avg}}[i]$ to get the following base-fee update rule
\begin{equation}\label{eq:mitigation}
    b[i+1] = b[i] \cdot \left( 1+\phi \cdot \frac{s_{\textit{avg}}[i]-\starget}{\starget}\right).
\end{equation}
We note that $b[0]$ would be initialized to 1 Gwei initially and that $s_{\textit{avg}}$ would be initialized to be the size of the first block after the transition. By applying the update rule in Equation~\ref{eq:mitigation}, we reduce the effect of the empty block on the ensuing block by a factor of~$(1-q)$ and discount its effects on later blocks exponentially fast (with base~$q$). For example, using the same parameters as before with $\phi=1/8$, if we set $q=1/2$, after $X$'s empty block (at slot~$\tau$), the base fee will be reduced by a factor of $(1-\phi(1-q))=(1-1/16)$ to $b[\tau+1]=15\btarget/16$. Now, however, making the same assumption as before (i.e., $s[\tau+1]=2\starget$), the base fee for the next block, $b[\tau+2]$, will be $495\btarget/512$. This decreases the potential profit margin of block $\tau+2$ from $\btarget/16$ to $\btarget/32 +\btarget/512$, which is almost a factor of~$2$ reduction. Therefore, the added profit opportunity in future blocks is not enough to compensate for the lost potential in the immediately ensuing block. As a consequence, the attack is considerably mitigated.

The above mitigation method has two additional properties: (i) its computation and space complexity are both in $O(1)$, and (ii) it gradually phases out the impact of a single empty block without causing significant fluctuations. To reason about the effect our proposal has on response times we use the following methodology. Suppose that the demand suddenly changes and the new (desired) steady state should be reached at a new point with a base fee that is $\beta$ times higher than the current base fee $(\btarget)$. Denote by $T$ the number of consecutive full blocks required to reach the new base fee. We use $T$ as a function of~$\beta$ to characterize the response time of a fee-setting mechanism to sudden changes in demand.

EIP-1559 as it is currently implemented will take $T(\beta)=\lceil\log_{(1+\phi)}\beta\rceil$ blocks to reach the new base fee~$\beta\btarget$. With our mitigation, it takes slightly longer. To be precise, in Appendix~\ref{app:mitigationDelay} we show the exact delay $T(\beta)$ of our mitigation proposal is the smallest integer $T$ that satisfies $\prod_{k=1}^T(1+\phi(1-q^k)) \ge \beta$. We further plot the $T(\beta)$ for EIP-1559 and our mitigation in Appendix~\ref{app:mitigationDelay} (cf. Figure~\ref{fig:time}).

\subsection{Simulations}
\begin{figure}[ht]
 \centering
\begin{subfigure}[t]{0.505\linewidth}
\centering
\includegraphics[scale =1]{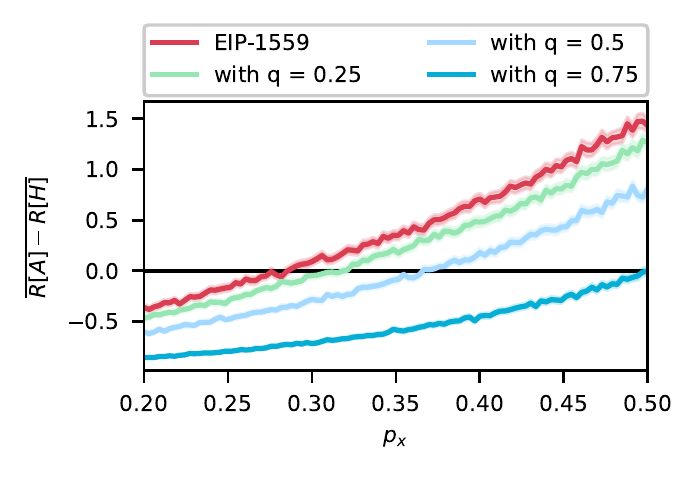}\vspace{-6pt}
\caption{Difference of mean attack and mean honest return as a function of $p_x$ for EIP-1559 and the proposed mitigation for $q\in\{1/4,1/2,3/4\}$. We set $\varepsilon/\btarget = 1/25$.} \label{fig:EIPvsMitigationPX}\vspace{-6pt}
\end{subfigure}   
\hfill
\begin{subfigure}[t]{0.475\linewidth}
\centering
\includegraphics[scale =1]{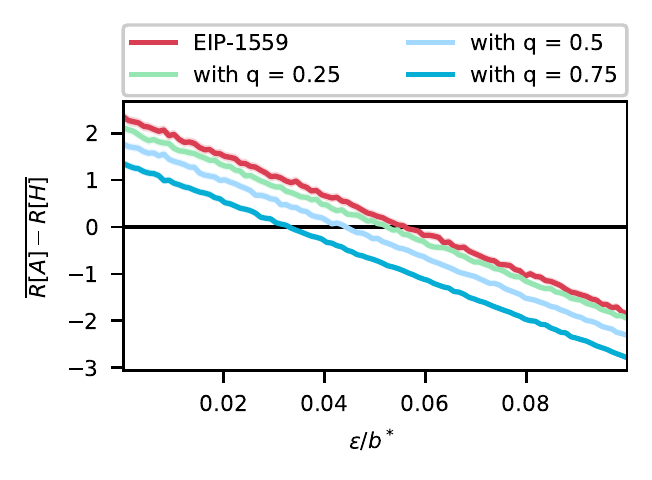}\vspace{-6pt}
\caption{Difference of mean attack and mean honest return as a function of $\varepsilon/\btarget$ for EIP-1559 and the proposed mitigation for $q\in\{1/4,1/2,3/4\}$. We set $p_x=0.4$.} \label{fig:EIPvsMitigationratiodb}\vspace{-6pt}
\end{subfigure}   
\caption{Profitability of the attack under EIP-1559 and our proposed mitigation. We plot the mean attack profitability (cf. Figures~\ref{fig:EIPvsMitigationPX} and~\ref{fig:EIPvsMitigationratiodb}) along with the 95\% confidence interval. We set $\phi =1/8$, $\alpha = 0.5$, $\starget = 1$.} \vspace{-6pt}\label{fig:mitigation}
\end{figure}

\begin{figure}[th]
 \centering
 \captionsetup[subfigure]{justification=centering}
\begin{subfigure}[t]{0.326\linewidth}
\centering
\includegraphics[scale =1]{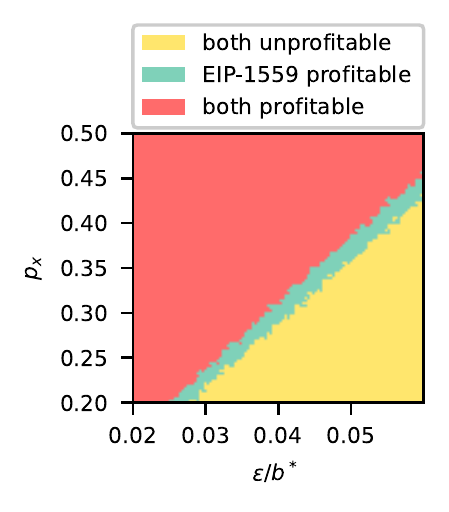}\vspace{-8pt}
\caption{ $q =1/4$} \label{fig:2Dbothq25}\vspace{-6pt}
\end{subfigure}   
\hfill
\begin{subfigure}[t]{0.326\linewidth}
\centering
\includegraphics[scale =1]{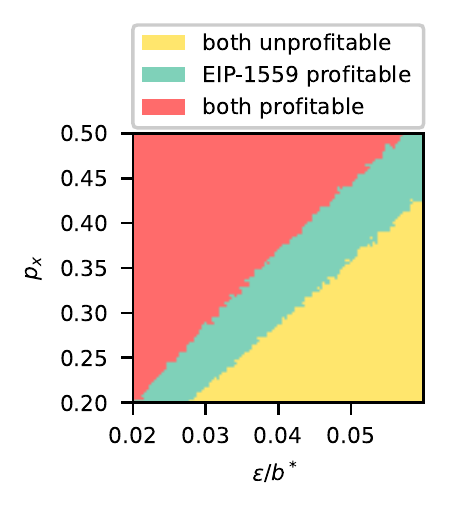}\vspace{-8pt}
\caption{ $q =1/2$}  \label{fig:2Dbothq50}\vspace{-6pt}
\end{subfigure}   
\hfill
\begin{subfigure}[t]{0.326\linewidth}
\centering
\includegraphics[scale =1]{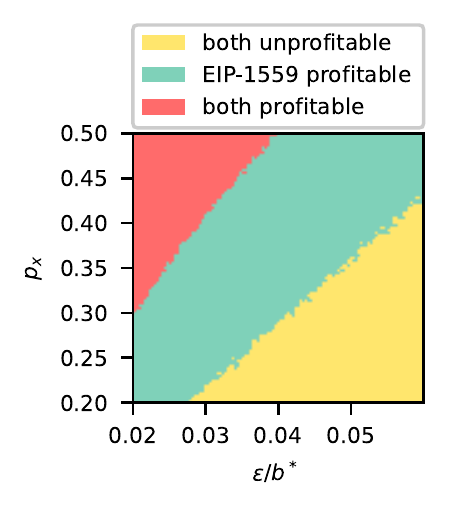}\vspace{-8pt}
\caption{ $q =3/4$} \label{fig:2Dbothq75}\vspace{-6pt}
\end{subfigure}  
\caption{Attack profitability under EIP-1559 and the proposed mitigation for $q\in\{1/4,1/2,3/4\}$ as a function of $p_x$ and $\varepsilon/\btarget$. Importantly, the green area shows where the mitigation can prevent the attack but EIP-1559 cannot. We set $\phi =1/8$, $\alpha = 0.5$, $\starget = 1$.} \vspace{-6pt}\label{fig:simulation}
\end{figure}

In order to gauge the effectiveness of the proposed mitigation, we conducted simulations that compare the excess profit of an attacker (profit gained through deviation minus profit gained through honest mining) under EIP-1559 with and without the mitigation.
To account for the probabilistic nature of the attack (profits in expectation only), we calculated the average of the results for each data point over 10,000 runs, each using a different random seed. Each simulation begins with $X$ mining an empty block, the next blocks are mined by $X$ with probability $p_x$ per block and by an honest miner with probability $1-p_x$. We assume that $X$ will then always mine target-size blocks, while the honest miners will mine blocks at twice the target size. We keep track of the payout received by $X$ and declare the attack finished if the base fee has recovered to 99\% of its target size. We perform 10,000 runs for both the base fee evolution according to EIP-1559, as well as our mitigation. Simultaneously, we keep track of the payout $X$ would have received for the same random seed if she had followed the honest strategy. The results of the simulation are depicted in Figures~\ref{fig:mitigation} and~\ref{fig:simulation}. 

Figures~\ref{fig:EIPvsMitigationPX} and~\ref{fig:EIPvsMitigationratiodb} show $X$'s profit from attacking as a function of $p_x$ and the ratio~$\varepsilon/\btarget$ for EIP-1559 with and without the mitigation for $q\in\{1/4, 1/2, 3/4\}$. The results decisively demonstrate the benefit of our mitigation; it becomes much harder for $X$ to profit from attacking. Finally, Figure~\ref{fig:simulation} illustrates the effect of the mitigation (with $q\in\{1/4,1/2,3/4\}$) on the configurations at which $X$ will attack. Most notably, the green area represents a set of configurations in which $X$ had attacked without the mitigation and will be attacking no longer. The value of the proposed approach is evident from the results.

\section{User perspective}~\label{sec:user}
Until now, we have approached the topic from the miners' point of view. Considering the users as first-class citizens, and observing the attack through their eyes, contributes a new perspective on the results. 

Instead of the miners taking it upon themselves to initiate the attack, we can imagine users who wish to pay lower costs coordinating the attack. Let~$u$ be a user (or group of users) that has transactions with a~$g$ amount of gas. Assume the other users naively follow the strategy of the desired equilibrium. That is, they bid an honest valuation with an~$\varepsilon$ tip. The attacker's strategy is as follows: $u$ bribes the miner of the current block (no matter the miner's power) to propose an empty one instead. Any bribe larger than~$\starget\varepsilon$ will suffice. Consequently, the base fee reduces in the next block. If the other users naively continue to bid with an~$\varepsilon$ tip (or they are simply slow to react), $u$ can guarantee the inclusion of her transactions with any tip larger than~$\varepsilon$ --- making the attack profitable whenever~$g\phi\btarget>\starget\varepsilon$.

\section{Discussion}~\label{sec:discussion}
\TA{Myopic vs. non-myopic.} 
Roughgarden's analysis~\cite{roughgarden2020transaction} of the EIP-1559 protocol suggests that, in the steady state, the honest strategies of the users and miners are incentive compatible. Our opposing conclusion stems from a single different assumption, namely, \cite{roughgarden2020transaction} considers miners that only care for immediate profits (referred to as myopic), while we consider non-myopic miners that do not disregard future profits. There are valid reasons to model miners as myopic. For example, the proposing turns of a very small miner are sporadic, and accounting for rare future profits is negligible in comparison to the prize at hand. However, there are strong reasons for modeling miners as non-myopic. Measurement studies have shown that the distribution of mining power follows a power law rather than a uniform distribution~\cite{gencer2018decentralization,grandjean2023ethereum}, i.e., the biggest miners control significant portions of the mining power. Moreover, the ``shorter vision'' of smaller miners is accounted for under our model. Our quantitative results show that as a miner's size decreases, the lesser the profitability of the deviations. Finally, from a conceptual perspective, miners are typically players of high stakes (the minimum stake in Ethereum is 32 ETH which is currently over \$50,000), and participation also requires some expertise, planning, and locking of assets. Therefore, it does not seem appropriate to consider these players as ones that neglect future considerations.

\T{Additional observations.}
The deviations described in this paper have an interesting property; they appear to have a win-win-win outcome. The attacker profits, the non-attacking miners profit, and the users profit by paying a cheaper total gas fee. But not all is rosy; there is a hidden cost involved. In order for users to benefit, they must diligently follow the miners' actions and compute the appropriate response. This increase in complexity for the users is in opposition to one of the main goals of EIP-1559 -- simplifying the bidding mechanism and eliminating the need for complex fee estimation algorithms. As a result, sophisticated users take precedence and push na\"ive users to the back of the line. 

Although it is desirable that the leader election process be unpredictable, in practice, this is not the case. Currently, in Ethereum, implementation considerations led to miners knowing their own proposing slots 32--62 blocks in advance~\cite{lookahead}. This predictability clearly favors the attacker, who no longer needs to lose tips for the probability of winning more. Instead, the miner only attacks when it is guaranteed to mine at least two blocks in a row. The predictability issue has risen with the move to PoS and was not present in PoW Ethereum.\footnote{Since the randomness in PoW does not depend on a peer-to-peer communication source, it does not have a predictability concern. The predictability issue is a result of implementation constraints for the cryptographic protocols of Verifiable Random Functions (VRFs).}

\section{Related work} \label{sec:related}

\T{Blockchain transaction fee mechanism.} Huberman et al.~\cite{huberman2021monopoly} provide an early analysis of Bitcoin's first-price auction. An in-depth exploration of miner manipulation in transaction fee mechanisms was first explored by Lavi et al.~\cite{lavi2022redesigning}. While these works study the first-price auction used in Bitcoin and originally in Ethereum as well, our work studies the incentives for miners to deviate from the EIP-1559 protocol currently used in Ethereum.

\T{Stability of the base fee.}
As Leonardos et al.~\cite{10.1145/3479722.3480993} show in their theoretical analysis, a key parameter in the base fee mechanism is the adjustment parameter $\phi$. In particular, they show that stability is not guaranteed depending on system conditions (e.g., a congested network), especially if $\phi$ is set to too high a value. However, there are also conditions under which the base fee may have bounded oscillations or even converge, providing stability. In another work, Leonardos et al.~\cite{leonardos2022optimality} show that despite the short-term chaotic behavior on the base fee, the long-term average block size is close to the target size.

Reijsbergen et al.~\cite{reijsbergen2021transaction} empirically show that a stable base fee may not tell the whole story, however, as even in cases where the base fee remains relatively stable (e.g., between 25 and 35 Gwei) and the block size is on average the target block size, block sizes can fluctuate wildly (as explained by Leonardos et al.~\cite{10.1145/3479722.3480993}), impacting miner revenue. One reason for this is that the currently used value for the adjustment parameter ($\phi=\frac{1}{8}$) is too low during periods where the demand rises sharply (i.e., the base fee does not increase quickly enough) but also too high when demand is stable (inducing fluctuations in block size). Therefore, \cite{reijsbergen2021transaction} suggests making $\phi$ variable based on the demand. Their work does not consider bribes and is complementary to ours. Their suggested mitigation to the stability issue is to have $\phi$ adaptive according to an Additive Increase Multiplicative Decrease (AIMD). Since we do not vary $\phi$ (but instead update the base fee update rule), an interesting experiment would be to combine our mitigation technique (averaging $s[i]$ geometrically) with their AIMD $\phi$ setting and examine the results on data from the real world. 

Ferreira et al.~\cite{ferreira2021dynamic} show that although the first-price auction utilized under EIP-1559 is incentive-aligned for miners, it provides a bad user experience. In particular, they observe bounded oscillation of the base fee in experiments when bidders all associate the same value with their transaction. While their work studies the stability of the base fee with myopic miners, we study the manipulability of the base fee in the presence of non-myopic miners.

\T{Manipulability of the base fee.}
Manipulation of the base fee has been a concern since EIP-1559 was proposed~\cite{eip1559outreach} as it is straightforward to notice that the base fee could be manipulated downwards by a miner that intentionally mines empty blocks.

The EIP as listed on Ethereum's Github repository acknowledges the possibility of miners mining empty blocks but determines that such a deviation from an honest mining strategy would not lead to a stable equilibrium as other miners would benefit from this (i.e., benefiting from the reduced base fee without the opportunity cost of mining an empty block)~\cite{eip1559}. Their belief was therefore that executing such a strategy would require a miner to control more than half the hashing power (the document precedes Ethereum's switch to PoS).

In his exposition of EIP-1559~\cite{roughgarden2020transaction}, Roughgarden considered this in the case of a 100\% miner (or any miner with greater than 50\% of the mining power) that would drive the base fee down to 0 by mining empty blocks then, in order to maximize their revenue, switching to either mining target size blocks in perpetuity, maintaining the base fee at 0 and essentially reverting back to a first price auction, or mining sequences of under full and overfull blocks (relative to the target size) according to the demand. Roughgarden restricts himself to this case and does not otherwise consider non-myopic miners, assuming that with a high enough level of decentralization, the probability of any miner being elected to propose a block two times in a row was too low for any miner to consider strategies over multiple blocks.

Similar to us, a concurrent work by Hougaard and Pourpouneh~\cite{hougaard2022farsighted} also relaxes the myopic assumption and proves that non-myopic miners would be incentivized to deviate even if they are a minority. However, their analysis relies on several assumptions, that while legitimate, provide the attacker with advantageous conditions in comparison to the more conservative assumptions made in this paper. In particular, \cite{hougaard2022farsighted}~relies on miners knowing the exact parameters of the demand distribution of the users (which itself is restricted to each user drawing from a uniform distribution), as well as on miners being able to manipulate the demand curve in favor of the future blocks; inducing artificial future congestion by not including transactions in the current block and letting them accumulate. We do not assume any specific demand curve, only a steady demand curve, and do not rely on miners inducing congestion, which makes our attack more profitable. Although we do assume that users act rationally and will adapt their strategy if benefits them, while~\cite{hougaard2022farsighted} assumes users are passive and do not adapt their strategies in a rational manner.

\T{MEV.}
Miner/Maximal Extractable Value (MEV) has gained significant attention in the blockchain research community in recent years~\cite{daian2020flash,torres2021frontrunner,qin2022quantifying}. Similarly to this work, MEV strategies enable a miner to accrue excess profits in comparison to na\"ive mining. However, in MEV the value comes from analyzing the actual data in the transactions, whereas, in this work, the miner's excess profit comes from manipulating the EIP-1559 mechanism. Therefore, many of the suggested mitigations against MEV~\cite{heimbach2022sok} (such as obscuring transaction data until inclusion) will not work against our attack.

\section{Conclusion and Future Work}\label{sec:conclusion}
In this paper, we demonstrated that even under very conservative assumptions (steady state, miners cannot induce congestion, unknown demand functions), there are strong incentives for minority miners to deviate from the EIP-1559 protocol. Furthermore, we showed that once an attack begins, previously honest miners' rational response may be to join the deviation and even sometimes initiate new attacks --- worsening the problem rather than improving it.

To mitigate the problem, we suggested using a weighted average with the weights being a geometric series. This direction seems promising as it has several desirable properties and trade-offs (e.g., balancing attack mitigation with low additional response delays). However, further research rigorously analyzing it in a broader context is warranted.

Many questions remain open regarding EIP-1559. To name but a few: While the suggested deviation captures the essence of the problem, devising an optimal deviation strategy is left as an open problem. What is the right abstraction to evaluate the benefits and shortcomings of EIP-1559 in comparison to a first-price auction? How does it differ from PoW to PoS? On a broader level, questions concerning adaptations of auction mechanisms for a blockchain use case have the potential to be both intellectually challenging and practically important and will hopefully receive more attention from the academic community.

\bibliography{ref}
\newpage
\appendix
\section{Omitted Proofs}\label{app:proofs}

\oneminer*

\begin{proof}

We commence with the honest strategy and calculate the expected payout for consecutive turns of $X$ as proposer by modeling the process as a Markov chain (cf. Figure~\ref{fig:markov_x_honest}).

When it is miner~$X$'s first time as proposer in a consecutive turn, i.e., the previous block was not mined by $X$, we are in state $X^h$. In this state, miner~$X$ proposes a block at target size $\starget$ and receives tips at price $\varepsilon$. Thus, the payout for miner~$X$ in state $X^h$ is
$
P[X^h]= \starget \cdot \varepsilon.
$
With probability $p_x$, $X$ stays in state $X^h$ for the next block and also proposes the next block.
Else, with probability $1-p_x$, we enter an absorbing state. We calculate the expected number of times $X$ proposes consecutive blocks and, thereby, the expected payout for $X$.

By linearity of expectation, it follows that the expected payout for a sequence of consecutive turns by $X$ as a proposer following the honest strategy is given by 
\begin{equation}
\mathbb{E}[R[X^h]] = p_x \cdot \mathbb{E}[ R[X^h]]  + P[X^h] \Longleftrightarrow \mathbb{E}[R[X^h]] =\frac{ \starget \cdot \varepsilon}{1-p_x},
\end{equation}
as miner~$X$ is awarded $ P[X^h]$ every time she proposes a block. Note that the expected reward of the honest strategy i.e., $\mathbb{E}[R[H]]$ corresponds to the expected payout of the Markov process starting in state $X^h$, i.e., $\mathbb{E}[R[H]] =\mathbb{E}[R[X^h]]$.
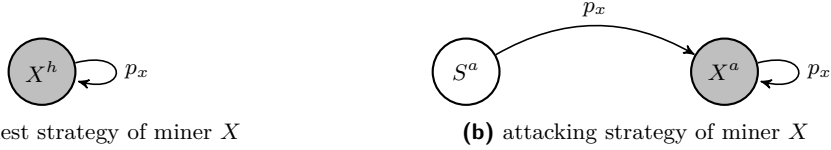
\begin{figure}[ht]
\centering
\captionsetup[subfigure]{justification=centering}
\begin{subfigure}[t]{0.38\linewidth}
\centering
\begin{tikzpicture}[,->, >=stealth', auto, semithick, node distance=2.5cm,font=\footnotesize,scale = 0.85]
\tikzstyle{every state}=[fill=white,draw=black,thick,text=black]
\node[state,fill = lightgray]    (x)  at (2.5,0)         {$X^h$};

\path
(x) edge[loop right]     node{$p_x$}     (x);
\end{tikzpicture}
\caption{honest strategy of miner~$X$} \label{fig:markov_x_honest}\vspace{-6pt}
\end{subfigure}
\hfill
\begin{subfigure}[t]{0.58\linewidth}
\centering
\begin{tikzpicture}[,->, >=stealth', auto, semithick, node distance=2.5cm,font=\footnotesize,scale = 0.85]

\tikzstyle{every state}=[fill=white,draw=black,thick,text=black]
\node[state]    (s)        at (0,0)              {$S^a$};
\node[state,fill = lightgray]    (x) at (4,0)       {$X^a$};
\path
(s) edge[bend left]     node{$p_x$}         (x)
(x) edge[loop right]     node{$p_x$}     (x)
;
\end{tikzpicture}
\caption{attacking strategy of miner~$X$} \label{fig:markov_x_attack}\vspace{-6pt}
\end{subfigure}   
\caption{The honest strategy (cf. Figure~\ref{fig:markov_x_honest}) and deviation from the honest strategy (cf. Figure~\ref{fig:markov_x_attack}) modeled with discrete Markov chains. All states with a nonzero payout for miner~$X$ are highlighted in gray. We transition between states with every block. Note that for all remaining probabilities, the Markov process enters an absorbing state, and $X$'s consecutive turns as a proposer finish.
}\label{fig:markov_x}\vspace{-6pt}
\end{figure}

We now examine the reward for miner~$X$ if she chooses to carry out the attack, modeling the deviation from the honest strategy as a Markov chain (as shown in Figure~\ref{fig:markov_x_attack}). The starting point for the attack, denoted as state $S^a$, is when miner~$X$ submits an empty block. Therefore, the payout in state $S^a$ is $0$. But with a probability of $p_x$, miner~$X$ is chosen as the proposer for the next block and enters state $X^a$, where she submits a block at the target size. The payout in state $X^a$ can be calculated by $P[X^a] = \starget \left( \phi\cdot \btarget +(1-\alpha)\varepsilon \right)$.

If $X$ is not selected to propose another block in a row, we enter an absorbing state and the attack stops. We make the approximation that whenever $X$'s consecutive turns as proposer finish, honest miners will bring the base fee back to $\btarget$ before $X$ gets to mine another block. It is possible that the honest miners do not bring the base fee back up to the target before $X$ is selected to propose again, and $X$ could continue the attack at a lower cost. However, assuming that the base fee had fully recovered simplifies the analysis and only makes our results stronger, as it reduces $X$'s attack rewards. From state $X^a$ we remain in state $X^a$ for the next block with probability $p_x$, i.e., $X$ is selected to propose another block or enter the absorbing state with probability $1-p_x$. 

The expected payout for miner~$X$ in state $S^a$ is given by 
$\mathbb{E}[R[S^a]] = p_x \cdot \mathbb{E}[R[X^a]] + P[S^a],$
where $\mathbb{E}[R [X^a]]$ is the expected payout starting from state $X^a$ and we have
$\mathbb{E}[R[X^a]] = p_x \cdot \mathbb{E}[R[X^a]] + P[X^a].$
It follows that the expected payout of the attack is 
\begin{equation}\mathbb{E}[R[A]] = \mathbb{E}[S^a] = \dfrac{p_x\cdot \starget \left(\phi\cdot \btarget +(1-\alpha)\varepsilon \right)}{(1-p_x)}.\end{equation}
Note that the payout of the attack is given by the expected payout starting from state $S^a$, as the attack commences in said state. 

We conclude that it is rational for miner~$X$ to deviate from the honest strategy if $\mathbb{E}[R[A]]  >\mathbb{E}[R[H]]$.
It follows that $X$ attacks when 
$$p_x > \tfrac{ \varepsilon}{\phi\cdot\btarget +(1-\alpha)\varepsilon}.$$
\end{proof}

\partialattack*
\begin{proof}
We commence with the honest strategy of miner~$Y$ which we model as a Markov chain in Figure~\ref{fig:markov_y_honest_partial}. Miner~$Y$ starts in state $Y^h_X$ and is tasked with proposing a block after miner~$X$ at an artificially lowered base fee $b$, where $b = (1-\phi) \btarget< \btarget$. The payout for proposing a block at twice the target size $\starget$ is given by
\begin{equation}P[Y^h_X] =\starget \left( \phi\cdot \btarget +(1-\alpha)\varepsilon \right) (1+\Delta).\end{equation} 
From state $Y^h_X$, we move to state $Y^h_Y$, where $Y$ proposes a target size block, with probability $p_y$. The payout for miner~$Y$ in state $Y^h_Y$ is given as 
\begin{equation}P[Y^h_Y] = \starget \cdot \varepsilon,\end{equation}
and we remain in this state for a subsequent block with probability $p_y$.

In both state $Y^h_X$ and state $Y^h_Y$, the probability of moving to state $X^h$ is $p_x$. When we re-enter state $X^h$ for the first time, $X$ will propose an empty block to lower the base fee again. Then, with a probability $p_x$, we remain in state $X$ for the next block, where miner~$X$ will propose target size blocks until her consecutive turn as a proposer is interrupted. Regardless, the payout for miner~$Y$ whenever we are in state $X^h$ is zero. We move to state $Y^h_Y$ with probability $p_y$ from state $X^h$ and enter an absorbing state with probability $1-p_y-p_x$ from all states. 

Next, we calculate the expected payout of the honest strategy and start with the expected reward for miner~$Y$ starting from state $Y^h_X$
$\mathbb{E}[R[Y^h_X]] =p_x \cdot \mathbb{E}[R[X^h]]+ p_y   \cdot \mathbb{E}[R[Y^h_Y]]+P[Y^h_X]$
where 
$\mathbb{E}[R[X^h]] = p_x \cdot \mathbb{E}[R[X^h]]+ p_y   \cdot \mathbb{E}[R[Y^h_X]],$
and 
$\mathbb{E}[R[Y^h_Y]] = p_x \cdot \mathbb{E}[R[X^h]]+ p_y   \cdot \mathbb{E}[R[Y^h_Y]]+P[Y^h_Y].$
By solving the system of linear equations, we find that the expected reward from the honest strategy is given by 
\begin{equation}
\begin{split}
    \mathbb{E}[R[H]] &= \mathbb{E}[R[Y^h_X]] &= \tfrac{(1-p_x)(((1-p_y)(1+\Delta)\phi \cdot \btarget)+  \varepsilon( 1+\Delta - \alpha (1+\Delta)(1-p_y)-\Delta p_y) )\starget}{(1-p_x - p_y)}.
\end{split}
\end{equation}

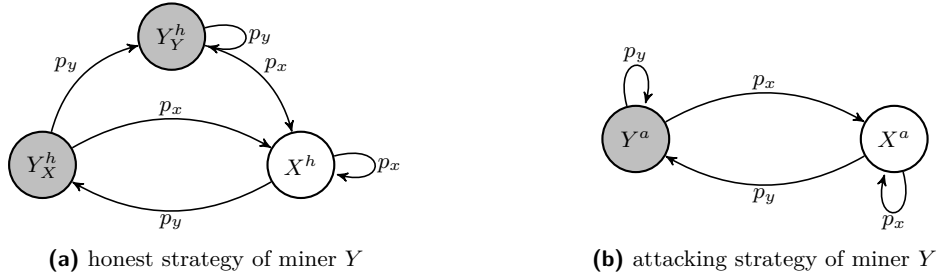
\begin{figure}[ht]
    \centering
    \captionsetup[subfigure]{justification=centering}
    \begin{subfigure}[t]{0.55\linewidth}
    \centering
    \begin{tikzpicture}[,->, >=stealth', auto, semithick, node distance=2.5cm,font=\footnotesize,scale = 0.85]
    \tikzstyle{every state}=[fill=white,draw=black,thick,text=black]
    \node[state]    (x) at (4,0)  {$X^h$};
    \node[state, fill = lightgray]    (xy) at (0,0)   {$Y^h_X$};
    \node[state, fill = lightgray]    (xyy) at (2,2)  {$Y^h_{Y}$};
    \path

    (x) edge[loop right]     node [inner sep=1pt] {$p_x$}         (x)
        edge[bend left]     node [inner sep=1pt] {$p_y$}         (xy)
    (xy) edge[bend left]     node [inner sep=1pt] {$p_x$}         (x)
        edge[bend left]     node [inner sep=1pt] {$p_y$}     (xyy)
    
    (xyy) edge[bend left]     node [inner sep=1pt] {$p_x$}         (x)
        edge[loop right]     node [inner sep=1pt] {$p_y$}     (xyy)
    ;
    \end{tikzpicture}
    \caption{honest strategy of miner~$Y$} \label{fig:markov_y_honest_partial}\vspace{-6pt}
    \end{subfigure}
    \hfill
    \begin{subfigure}[t]{0.4\linewidth}
    \centering
    \begin{tikzpicture}[,->, >=stealth', auto, semithick, node distance=2.5cm,font=\footnotesize,scale = 0.85]
    
    \tikzstyle{every state}=[fill=white,draw=black,thick,text=black]
    \node[state, fill = lightgray]    (y)       at (0,0)   {$Y^a$};
    \node[state]    (x) at (4,0) {$X^a$};
    \path

    (x) edge[loop below]     node [inner sep=1pt]{$p_x$}         (x)
        edge[bend left]     node [inner sep=1pt] {$p_y$}         (y)
    (y) edge[loop above]     node  [inner sep=1pt] {$p_y$}         (y)
        edge[bend left]     node [inner sep=1pt] {$p_x$}         (x)
    ;
    \end{tikzpicture}
    \caption{attacking strategy of miner~$Y$} \label{fig:markov_y_attack_partial}\vspace{-6pt}
    \end{subfigure}   
    \caption{The honest strategy (cf. Figure~\ref{fig:markov_y_honest_partial}) and the deviation from the honest strategy, i.e., keep the base fee artificially low, (cf. Figure~\ref{fig:markov_y_attack_partial}) modeled with discrete Markov chains for miner~$Y$. We transition between states with every block. All states with a nonzero payout for miner~$Y$ are highlighted in gray. Note that for all remaining probabilities, the Markov process enters an absorbing state.
    }\label{fig:markov_y_partial}\vspace{-6pt}
\end{figure}

We now consider the deviation from the honest strategy, whereby miner~$Y$ also keeps the base fee artificially low, and model the strategy with a Markov chain in Figure~\ref{fig:markov_y_attack_partial}. State $Y^a$ is the starting state of the deviating strategy in which $Y$ proposes a block with target size $\starget$ at an artificially lowered base fee $(1-\phi) \btarget$. Thus, the state's payout is 
\begin{equation}
P[Y^a] = \starget \left(\phi \cdot \btarget+(1-\alpha)\varepsilon \right).
\end{equation}

For the next block, we stay in state $Y^a$ with probability $p_y$, move to state $X^a$ with probability $p_x$, or else move to an absorbing state, i.e., the consecutive turns of $X$ and $Y$ as proposers end and the attack finishes. In state $X^a$, proposer $X$ will also propose a target size block, but the payout for miner~$Y$ is zero as we assume no collaboration between the two. The transition probabilities from state  $X^a$ are identical to those from state $Y^a$: we move to state $Y^a$ with probability $p_y$, stay in state $X^a$ with probability $p_x$ and enter an absorbing state with probability $1-p_y-p_x$ for the next block. Thus, the expected reward for miner~$Y$ starting in state $Y^a$ is given by 
$
\mathbb{E}[R[Y^a]] = p_x \cdot \mathbb{E}[R[X^a]] + p_y \cdot \mathbb{E}[R[Y^a]] +  P[Y^a],
$
where $\mathbb{E}[R[X^a]]$ is the expected payout for miner~$Y$ starting from state $X^a$ and we have 
$
\mathbb{E}[R[X^a]] = p_x \cdot \mathbb{E}[R[X^a]] + p_y \cdot \mathbb{E}[R[Y^a]].
$
We follow that the expected payout of the deviating strategy for miner~$Y$ is given by
\begin{equation}
\mathbb{E}[R[A]] = \mathbb{E}[R[Y^a]] = \dfrac{(1-p_x)\starget(\phi\cdot \btarget+ (1-\alpha)\varepsilon)}{ (1-p_x-p_y)}.
\end{equation}
We conclude it is rational for miner~$Y$ to deviate from the honest strategy when $\mathbb{E}[R[A]]  -\mathbb{E}[R[H]]>0$. It follows that $Y$ attacks when 
$$
p_y >\tfrac{\Delta ((1-\alpha)\varepsilon + \phi\cdot \btarget)}{(1-\alpha)\Delta\cdot\varepsilon +(1+\Delta) \phi\cdot \btarget - \alpha\cdot\varepsilon}
.
$$
\end{proof}
\entireattack*
\begin{proof}
We, again, model the honest strategy for miner~$Y$ as a Markov chain (cf. Figure~\ref{fig:markov_y_honest}). The honest strategy starts in state $Y^h$, where miner~$Y$ proposes a block at target size $\starget$ and receives a payout of 
\begin{equation}
P[Y^h] = \starget \cdot \varepsilon.
\end{equation} 
The transition probability to state $X^h$ is $p_x$ and to state $Y^h$, i.e., $Y$ proposes consecutive blocks, is $p_y$. Otherwise, the consecutive turn of $X$ and $Y$ as miners stops and we enter an absorbing state. We note that the states $X^h$, $Y_X^h$ and $Y_Y^h$ correspond exactly to the eponymous states in Theorem~\ref{thm:partialattack} (cf. Figure~\ref{fig:markov_y_honest_partial}). Thus, the actions of the miners, the payout for miner~$Y$, and the transition probabilities are as previously described in the proof of Theorem~\ref{thm:partialattack}. 

The strategy's payout is the expected reward starting from state $Y^h$, which is
\begin{equation}
\mathbb{E}[R[Y^h]] =p_x \cdot \mathbb{E}[R[X^h]]+ p_y   \cdot \mathbb{E}[R[Y^h]+P[Y^h].
\end{equation}
As previously in Theorem~\ref{thm:partialattack}, we have
\begin{align}
    \mathbb{E}[R[Y^h_X]] &=p_x \cdot \mathbb{E}[R[X^h]]+ p_y   \cdot \mathbb{E}[R[Y^h_Y]]+P[Y^h_X],\\
    \mathbb{E}[R[X^h]] &= p_x \cdot \mathbb{E}[R[X^h]]+ p_y   \cdot \mathbb{E}[R[Y^h_X]],\\
    \mathbb{E}[R[Y^h_Y]] &= p_x \cdot \mathbb{E}[R[X^h]]+ p_y   \cdot \mathbb{E}[R[Y^h_Y]]+P[Y^h_Y].
\end{align}
Solving the system of linear equation, we conclude that the expected payout of the honest strategy is
\begin{equation}
\mathbb{E}[R[H]] = \mathbb{E}[R[Y^h]] = \frac{(((1+\Delta)\btarget \phi p_xp_y) + \varepsilon(1-p_x + ((1-\alpha)\Delta -\alpha)p_xp_y))\starget}{1-p_x -p_y}.
\end{equation}
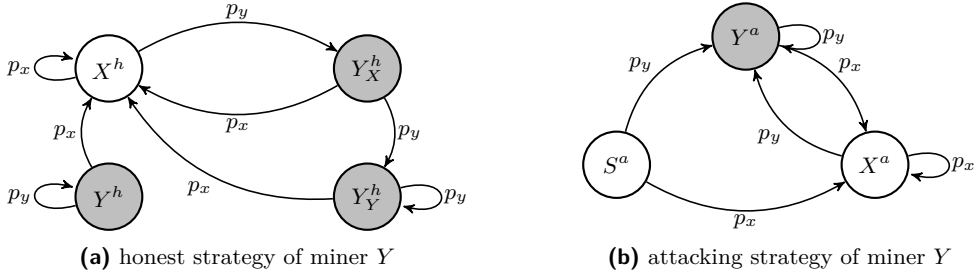
\begin{figure}[ht]
\centering
\captionsetup[subfigure]{justification=centering}
\begin{subfigure}[t]{0.49\linewidth}
\centering
\begin{tikzpicture}[,->, >=stealth', auto, semithick, node distance=2.5cm,font=\footnotesize,scale = 0.85]
\tikzstyle{every state}=[fill=white,draw=black,thick,text=black]
\node[state, fill = lightgray]    (y)  at (4,0)       {$Y^h$};
\node[state]    (x) at (4,2)  {$X^h$};
\node[state, fill = lightgray]    (xy) at (8,2)   {$Y^h_X$};
\node[state, fill = lightgray]    (xyy) at (8,0)  {$Y^h_{Y}$};
\path
(y) edge[bend left]     node [inner sep=1pt] {$p_x$}         (x)
    edge[loop left]     node [inner sep=1pt] {$p_y$}     (y)
(x) edge[loop left]     node [inner sep=1pt] {$p_x$}         (x)
    edge[bend left]     node [inner sep=1pt] {$p_y$}         (xy)
(xy) edge[bend left]     node [inner sep=1pt] {$p_x$}         (x)
    edge[bend left]     node [inner sep=1pt] {$p_y$}     (xyy)

(xyy) edge[bend left]     node [inner sep=1pt] {$p_x$}         (x)
    edge[loop right]     node [inner sep=1pt] {$p_y$}     (xyy)
;
\end{tikzpicture}
\caption{honest strategy of miner~$Y$} \label{fig:markov_y_honest}\vspace{-6pt}
\end{subfigure}
\hfill
\begin{subfigure}[t]{0.49\linewidth}
\centering
\begin{tikzpicture}[,->, >=stealth', auto, semithick, node distance=2.5cm,font=\footnotesize,scale = 0.85]
    
    \tikzstyle{every state}=[fill=white,draw=black,thick,text=black]
    \node[state]    (s)       at (0,0)   {$S^a$};
    \node[state]    (x) at (4,0) {$X^a$};
    \node[state, fill = lightgray]    (y) at (2,2) {$Y^a$};
    \path
    (s) edge[bend right]     node [inner sep=1pt,below]{$p_x$}         (x)
        edge[bend left]     node [inner sep=1pt] {$p_y$}         (y)
    
    (x) edge[loop right]     node [inner sep=1pt]{$p_x$}         (x)
        edge[bend left]     node [inner sep=1pt] {$p_y$}         (y)
    (y) edge[loop right]     node  [inner sep=1pt] {$p_y$}         (y)
        edge[bend left]     node [inner sep=1pt] {$p_x$}         (x)
    ;
\end{tikzpicture}
\caption{attacking strategy of miner~$Y$} \label{fig:markov_y_attack}\vspace{-6pt}
\end{subfigure}   
\caption{The honest strategy (cf. Figure~\ref{fig:markov_y_honest}) and the deviation from the honest strategy (cf. Figure~\ref{fig:markov_y_attack}) modeled with discrete Markov chains. We transition between states with every block. All states with a nonzero payout for miner~$Y$ are highlighted in gray. Note that for all remaining probabilities, the Markov process enters an absorbing state and the consecutive turn of $X$ and $Y$ as proposers finishes.
}\label{fig:markov_y_full}\vspace{-6pt}
\end{figure}

We proceed with the attack strategy, which we model in Figure~\ref{fig:markov_y_attack}. Miner~$Y$ starts in state $S^a$ and mines an empty block and, therefore, receives no rewards. With probability $p_y$ we move to state $Y^a$ for the next block, i.e., $Y$ proposes a target size ($\starget$) block, with probability $p_x$ we move to state $X^a$, i.e., $X$ proposes a target size ($\starget$) block, and with probability $1-p_x-p_y$ we move to an absorbing state, i.e., the attack ends. Notice that the states $X^a$ and $Y^a$ are identical to those described in the proof of Theorem~\ref{thm:entireattack}. Thus, the expected returns starting in the respective states are as follows
\begin{align}
    \mathbb{E}[R[S^a]] &= p_x \cdot \mathbb{E}[R[X^a]] + p_y \cdot \mathbb{E}[R[Y^a]],\\
    \mathbb{E}[R[Y^a]] &= p_x \cdot \mathbb{E}[R[X^a]] + p_y \cdot \mathbb{E}[R[Y^a]] +  P[Y^a],\\
    \mathbb{E}[R[X^a]] &= p_x \cdot \mathbb{E}[R[X^a]] + p_y \cdot \mathbb{E}[R[Y^a]],
\end{align}
and find that the expected reward of the attack strategy is 
\begin{equation}
\mathbb{E}[R[A]] = \mathbb{E}[R[S^a]]= \frac{(\phi\cdot \btarget + (1 - \alpha) \varepsilon)p_y\starget}{ 1-p_x-p_y}.
\end{equation}
To conclude, it is rational behavior for $Y$ to deviate from the honest strategy, when $\mathbb{E}[R[A]]-\mathbb{E}[R[H]]> 0$, which holds when 
$$p_y > \frac{\varepsilon (1-p_x)}{(1-\alpha) \varepsilon + \phi \cdot \btarget (1-p_x) + (1+\Delta)\varepsilon\alpha p_x - \Delta p_x(\varepsilon+\phi\cdot \btarget)}.
$$
\end{proof}
\section{Delay Incurred by the Mitigation of Section~\ref{sec:mitigation}}
\label{app:mitigationDelay}

Suppose we are in a steady state with a base fee $\btarget$, when after block height~$\tau$ a fixed change in demand occurs for which the new steady state will be achieved with the new base fee $\beta\cdot\btarget$. We denote by~$T$ the number of consecutively full blocks it takes to reach the new base fee $\beta\cdot\btarget$.

After~$k$ consecutively full blocks, according to Eq.~\ref{eq:Savg}
\begin{align*}
    s_{\textit{avg}}[\tau+k] &= (1-q)2\starget+q\cdot s_{\textit{avg}}[\tau+k-1]\\
    &= \left(2(1-q)(q^0+q^1+\ldots+q^{k-1})+q^k\right)\starget \\
    &= \left( 2(1-q)\left( \frac{1-q^k}{1-q} \right) +q^k \right)\starget \\
    &= (2-q^k)\starget. 
\end{align*}
Plugging the above into Eq.~\ref{eq:mitigation} yields
\begin{align*}
    b[\tau+k] &= b_{\textit{avg}}[\tau+k-1] \cdot \left(1+\phi\cdot \frac{(2-q^k)\starget-\starget}{\starget}\right) = b_{\textit{avg}}[\tau+k-1] \cdot \left(1+\phi (1-q^k) \right) \\
    &= b[\tau]\cdot \left(1+\phi (1-q^1) \right) \left(1+\phi (1-q^2) \right) \cdots \left(1+\phi (1-q^k) \right) \\
    &= \btarget \cdot \prod_{i=1}^k(1+\phi(1-q^i)).
\end{align*}
 Therefore, $T$ is the smallest integer that satisfies 
$$
b[\tau+T] = \btarget \cdot \prod_{i=k}^T(1+\phi(1-q^k)) \ge \beta\cdot\btarget \quad
\Longleftrightarrow \quad
\prod_{k=1}^T(1+\phi(1-q^k)) \ge \beta.
$$ 

\begin{figure}[ht]
\centering
\includegraphics[scale=1]{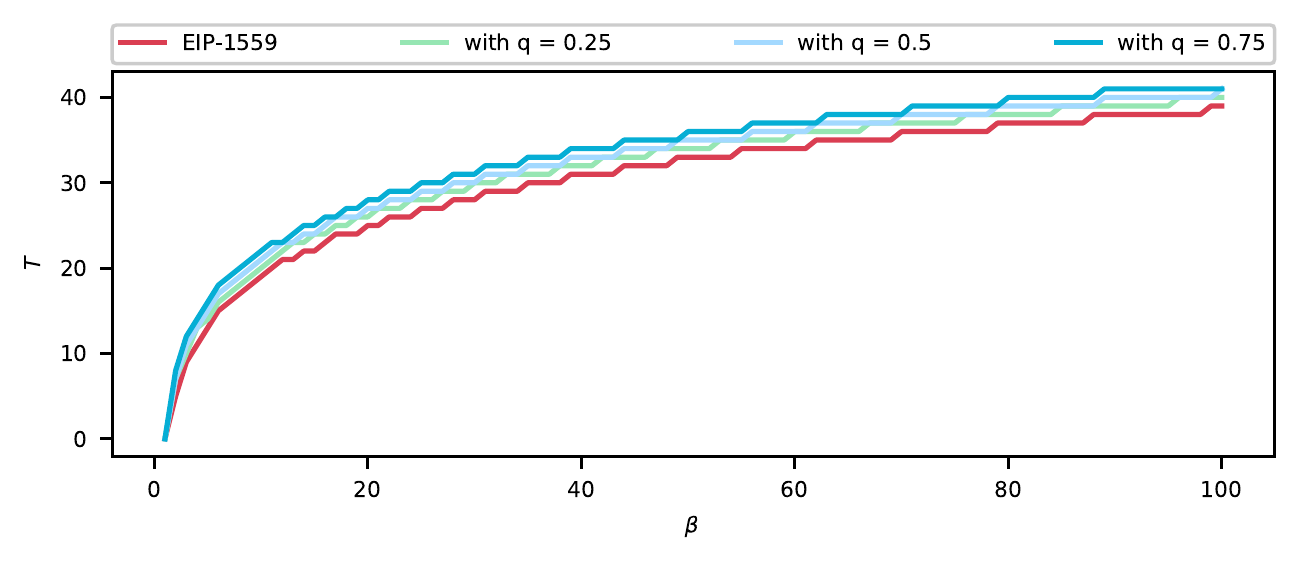}\vspace{-8pt}
   \caption{The number of consecutive full blocks $T$ required to increase the base fee by factor $\beta$ for EIP-1559 and the proposed mitigation for $q\in\{1/4,1/2,3/4\}$. }\label{fig:time}
    \vspace{-8pt}
\end{figure}

Figure~\ref{fig:time} plots $T(\beta)$ for both EIP-1559 and our mitigation. We set~$\phi=1/8$ (current Ethereum) and use $q\in\{1/4, 1/2, 3/4\}$. All plots follow a logarithmic trend and the response times to dramatic changes in demand are only mildly affected by the proposed mitigation (even for~$\beta$ factors as large as 100).

\end{document}